\documentclass{article}

\usepackage[preprint, nonatbib]{neurips_2019}

\usepackage{amsmath,amsfonts,bm}

\def\eqref#1{equation~\ref{#1}}
\def\1{\bm{1}}

\def\mA{{\bm{A}}}

\def\mD{{\bm{D}}}

\def\mH{{\bm{H}}}
\def\mI{{\bm{I}}}

\def\mW{{\bm{W}}}
\def\mX{{\bm{X}}}

\DeclareMathAlphabet{\mathsfit}{\encodingdefault}{\sfdefault}{m}{sl}
\SetMathAlphabet{\mathsfit}{bold}{\encodingdefault}{\sfdefault}{bx}{n}

\def\gG{{\mathcal{G}}}

\def\sR{{\mathbb{R}}}

\usepackage{hyperref}
\usepackage{url}

\usepackage[utf8]{inputenc} % allow utf-8 input
\usepackage[T1]{fontenc}    % use 8-bit T1 fonts
\usepackage{comment}
\usepackage{hyperref}       % hyperlinks
\usepackage{url}            % simple URL typesetting
\usepackage{graphicx}% Include figure files
\usepackage{booktabs}       % professional-quality tables
\usepackage{amsfonts}       % blackboard math symbols
\usepackage{nicefrac}       % compact symbols for 1/2, etc.
\usepackage{microtype}      % microtypography
\usepackage{wrapfig}
\usepackage{bm}
\usepackage{braket}
\usepackage{amsmath,amssymb,latexsym}
\usepackage{subcaption}

\newcommand{\droppedfootnote}[1]{}

\DeclareMathOperator{\relu}{ReLU}
\DeclareMathOperator{\ks}{KS}

\title{Quantum Graph Neural Networks}

\author{
  Guillaume~Verdon \\
  X, The Moonshot Factory \\
  Mountain View, CA\\
  \texttt{gverdon@x.team} \\
   \And
   Trevor~McCourt \\
  Google Research\\
  Venice, CA\\
  \texttt{trevormccrt@google.com} \\
   \AND
   Enxhell~Luzhnica, Vikash~Singh,\\ \textbf{Stefan~Leichenauer, Jack~Hidary }\\
  X, The Moonshot Factory\\
  Mountain View, CA \\
  \texttt{\{enxhell,singvikash,}\\\texttt{sleichenauer,hidary\}@x.team} \\
}
\begin{document}

\maketitle

\begin{abstract}
We introduce Quantum Graph Neural Networks (\textsc{qgnn}), a new class of quantum neural network ansatze which are tailored to represent quantum processes which have a graph structure, and are particularly suitable to be executed on distributed quantum systems over a quantum network. Along with this general class of ansatze, we introduce further specialized architectures, namely, Quantum Graph Recurrent Neural Networks (\textsc{qgrnn}) and Quantum Graph Convolutional Neural Networks (\textsc{qgcnn}). We provide four example applications of \textsc{qgnn}s: learning Hamiltonian dynamics of quantum systems, learning how to create multipartite entanglement in a quantum network, unsupervised learning for spectral clustering, and supervised learning for graph isomorphism classification. 

\end{abstract}

\section{Introduction}

Variational Quantum Algorithms are a promising class of algorithms that are rapidly emerging as a central subfield of Quantum Computing \cite{mcclean2016theory, farhi2014quantum, farhi2018classification}. Similar to parameterized transformations encountered in deep learning, these parameterized quantum circuits are often referred to as Quantum Neural Networks (QNNs). Recently, it was shown that QNNs that have no prior on their structure suffer from a quantum version of the no-free lunch theorem~\cite{mcclean2018barren} and are exponentially difficult to train via gradient descent. Thus, there is a need for better QNN ansatze. One popular class of QNNs has been Trotter-based ansatze \cite{farhi2014quantum, hadfield2019quantum}. The optimization of these ansatze has been extensively studied in recent works, and efficient optimization methods have been found~\cite{verdon2019learning, li2019quantum}. On the classical side, graph-based neural networks leveraging data geometry have seen some recent successes in deep learning, finding applications in biophysics and chemistry~\cite{kearnes2016molecular}. Inspired from this success, we propose a new class of Quantum Neural Network ansatz which allows for both quantum inference and classical probabilistic inference for data with a graph-geometric structure. In the sections below, we introduce the general framework of the \textsc{qgnn} ansatz as well as several more specialized variants and showcase four potential applications via numerical implementation.

\begin{comment}
\begin{itemize}
\item[-] Many quantum processing chips have restricted or non-grid topologies
\item[ -] Need for quantum neural network ansatz which takes into account the communication overhead and fits directly on the topology of the network or chip
\item[-] In the case of quantum chips, two-qubit gates remain the fidelity bottleneck, while in the case of quantum networks, two-node operations require rounds of inter-node communication.
\item[-] Graph neural networks have seen a large growth in the number of areas of application on the classical side.
\item[-] Generally these have degrees of freedom on both the nodes and edges of the network
\item[-] In our case, the restriction to networks with quantum degrees of freedom on the nodes only reduces the overall volume of quantum memory needed. 
\item[-] By having the edges be classsical and the nodes quantum, we open up the possibilty of extracting insights about various quantum systems of interest without having to simulate an entire volume of space as a grid.
\end{itemize}
\end{comment}

%In this section we define the general form of the Quantum Graph Neural Network ansatz, as well as the ``convolutional'' and ``recurrent'' specialized subtypes.

\section{Background}
\subsection{Classical Graph Neural Networks}
Graph Neural Networks (\textsc{gnn}s) date back to \cite{sperduti1997supervised} who applied neural networks to acyclic graphs. \cite{gori2005new} and \cite{scarselli2008graph} developed methods that learned node representations by propagating the information of neighbouring nodes. Recently, \textsc{gnn}s have seen great breakthroughs by adapting the convolution operator from \textsc{cnn}s to graphs \cite{bruna2013spectral,henaff2015deep,defferrard2016convolutional,kipf2016semi,niepert2016learning,hamilton2017inductive,monti2017geometric}. Many of these methods can be expressed under the message-passing framework \cite{gilmer2017neural}.

Let graph $\gG=(\mA,\mX)$ where $\mA \in \sR^{n \times n}$ is the adjacency matrix, and $\mX \in \sR^{n \times d}$ is the node feature matrix where each node has $d$ features. 

\begin{equation}
	\mH^{(k)} = P(\mA, \mH^{(k-1)}, \mW^{(k)})
\end{equation}

where $\mH^{(k)} \in \sR^{n \times d}$ are the node representations computed at layer $k$, $P$ is the message propagation function and is dependent on the adjacency matrix, the previous node encodings and some learnable parameters $\mW^{(k)}$. The initial embedding, $\mH^{(0)}$, is naturally $\mX$. One popular implementation of this framework is the \textsc{gcn} \cite{kipf2016semi} which implements it as follows: %should all mentions of NN's be in textsc?

\begin{equation}
	\mH^{(k)} = P(\mA,\mH^{(k-1)}, \mW^{(k)}) = \relu(\Tilde{\mD}^{-\frac{1}{2}}\Tilde{\mA}\Tilde{\mD}^{-\frac{1}{2}}\mH^{(k-1)}\mW^{(k-1)})
\end{equation}

where $\Tilde{\mA} = \mA + \mI$ is the adjacency matrix with inserted self-loops and $\Tilde{\mD} = \sum_{j} \Tilde{\mA}_{ij}$ is the renormalization factor (degree matrix).

\subsection{Networked Quantum Systems} Consider a graph $\mathcal{G} = \{\mathcal{V},\mathcal{E}\}$, where $\mathcal{V}$ is the set of vertices (or nodes) and $\mathcal{E}$ the set of edges. We can assign a quantum subsystem with Hilbert space $\mathcal{H}_v$ to each vertex in the graph, forming a global Hilbert space $\mathcal{H}_\mathcal{V} \equiv \bigotimes_{v\in\mathcal{V}} \mathcal{H}_v$. Each of the vertex subsystems could be one or several qubits, a qudit, a qumode \cite{weedbrook2012gaussian}, or even an entire quantum computer. {One may also define a Hilbert space for each edge and form $\mathcal{H}_\mathcal{E} \equiv \bigotimes_{e\in\mathcal{E}} \mathcal{H}_e$. The total Hilbert space for the graph would then be $\mathcal{H}_\mathcal{E}\otimes \mathcal{H}_\mathcal{V}$. For the sake of simplicity and feasibility of numerical implementation, we consider this to be beyond the scope of the present work, so for us the total Hilbert space consists only of $\mathcal{H}_\mathcal{V}$.} The edges of the graph dictate the communication between the vertex subspaces: couplings between degrees of freedom on two different vertices are allowed if there is an edge connecting them. This setup is called a quantum network \cite{kimble2008quantum,qian2019heisenberg} with topology given by the graph $\mathcal{G}$.

%Quantum algorithms on a network can be considered as a sequence of operations which involve only neighboring nodes at a time.
\section{Quantum Graph Neural Networks}

\subsection{General Quantum Graph Neural Network Ansatz} The most general Quantum Graph Neural Network ansatz is a parameterized quantum circuit on a network which consists of a sequence of $Q$ different Hamiltonian evolutions, with the whole sequence repeated $P$ times:
\begin{equation}\label{eq:QGNN}
    \hat{U}_{\textsc{qgnn}}(\bm{\eta}, \bm{\theta}) = \prod_{p=1}^P \left[ \prod_{q=1}^Q e^{-i\eta_{pq}\hat{H}_q(\bm{\theta})}\right],
\end{equation}
where the product is time-ordered \cite{poulin2011quantum}, the $\bm{\eta}$ and $\bm{\theta}$ are variational (trainable) parameters, and the Hamiltonians $\hat{H}_q(\bm{\theta})$ can generally be any parameterized Hamiltonians whose topology of interactions is that of the problem graph:
\begin{equation}\label{eq:Ham} \hat{H}_q(\bm{\theta}) \equiv \sum_{\{j,k\}\in \mathcal{E}} \sum_{r\in\mathcal{I}_{jk}} W_{qrjk} \hat{O}^{(qr)}_{j}\otimes \hat{P}^{(qr)}_{k}+ \sum_{v\in \mathcal{V}} \sum_{r\in\mathcal{J}_{v}} B_{qrv} \hat{R}^{(qv)}_{j}.\end{equation}
Here the $W_{qrjk}$ and $B_{qrv}$ are real-valued coefficients which can generally be independent trainable parameters, forming a collection $\bm{\theta} \equiv \cup_{q,j,k,r}\{W_{qrjk}\}\cup_{q,v,r}\{B_{qrjk}\}$. The operators $\hat{R}^{(qv)}_{j}, \hat{O}^{(qr)}_{j}, \hat{P}^{(qr)}_{j}$ are Hermitian operators which act on the Hilbert space of the $j^\mathrm{th}$ node of the graph. The sets $\mathcal{I}_{jk}$ and $\mathcal{J}_{v}$ are index sets for the terms corresponding to the edges and nodes, respectively. To make compilation easier, we enforce that the terms of a given Hamiltonian $\hat{H}_q$ commute with one another, but different $\hat{H}_q$'s need not commute.%\footnote{Note that the operator sum decomposition can be expressed equivalently by various choices of basis for the space of operators (e.g., the $n$-qubit Paulis).} Despite this restriction, the ansatz structure in \eqref{eq:QGNN} is sufficiently general to cover the any case where one has a parametrized quantum circuit ansatz whose network topology is that of the graph $\mathcal{G}$.  %As an example, in the case where we assign several qubits per node, in this case, assuming we assign $n_j$ qubits per node, one choice of basis which spans the space of operators is the set of of $n_j$-qubit Pauli's $\mathcal{P}_{n_j}$.

%For the \textsc{qgnn}, we enforce that the terms of a given Hamiltonian $\hat{H}_q$ commute with one another, but generally different $\hat{H}_q$'s do not commute with one another. The reason for this enforcement is to avoid for the need of Trotterization in the compilation of the exponential \cite{}. 

In order to make the ansatz more amenable to training and avoid the barren plateaus (quantum parametric circuit no free lunch) problem~\cite{mcclean2018barren}, we need to add some constraints and specificity. To that end, we now propose more specialized architectures where parameters are tied spatially (convolutional) or tied over the sequential iterations of the exponential mapping (recurrent).

\subsection{Quantum Graph Recurrent Neural Networks (\textsc{qgrnn})} We define quantum graph recurrent neural networks as ansatze of the form of \eqref{eq:QGNN} where the temporal parameters are tied between iterations, $\eta_{pq} \mapsto \eta_q$. In other words, we have tied the parameters between iterations of the outer sequence index (over $p=1,\ldots, P$). This is akin to classical recurrent neural networks where parameters are shared over sequential applications of the recurrent neural network map. As $\eta_q$ acts as a time parameter for Hamiltonian evolution under $\hat{H}_q$, we can view the \textsc{qgrnn} ansatz as a Trotter-based \cite{lloyd1996universal,poulin2011quantum} quantum simulation of an evolution $e^{-i\Delta \hat{H}_{\mathrm{eff}}}$ under the Hamiltionian 
\(\hat{H}_{\mathrm{eff}} = \Delta^{-1}\sum_q \eta_q\hat{H}_q\) for a time step of size $\Delta =\lVert\bm{\eta}\rVert_1 = \sum_q |\eta_q|$. This ansatz is thus specialized to learn effective quantum Hamiltonian dynamics for systems living on a graph. In Section~\ref{sec:dynamics} we demonstrate this by learning the effective real-time dynamics of an Ising model on a graph using a \textsc{qgrnn} ansatz.

% %%%%%commented out
% \begin{comment}
% \[U_{\textsc{qgrnn}}(\bm{\eta}, \bm{\theta}) = \prod_{l=1}^P \left[ \prod_{q=1}^Q e^{-i\eta_{q}\hat{H}_q(\bm{\theta}_q)}\right]      \]
% where the $\bm{\eta}$ parameters are variational parameters, and the parameterized Hamiltonians $\hat{H}_q(\bm{\theta}_q)$ can generally be any 2-local Hamiltonians whose topology of interactions is that of the problem graph:

% \begin{equation}\label{eq:Ham} \hat{H}_q(\bm{\theta}_q) \equiv \sum_{\{j,k\}\in \mathcal{E}} \sum_{r\in\mathcal{I}_{jk}} W_{qrjk} \hat{O}^{(qr)}_{j}\otimes \hat{P}^{(qr)}_{k}+ \sum_{v\in \mathcal{V}} \sum_{r\in\mathcal{J}_{v}} B_{qrv} \hat{R}^{(qv)}_{j}.\end{equation}

% Here, we labelled the parameters $\bm{\theta} \equiv \cup_{q,j,k,r}\{W_{qrjk}\}\cup_{q,v,r}\{B_{qrjk}\}$, which are each \textit{independently} variational (i.e. trainable).

% Importantly, notice that the parameters $\eta_{q}$ are independent of the layer index $l$. In other words, we have tied the parameters between iterations of the outer sequence index, this is akin to classical recurrent neural networks where parameters are shared over sequential applications of the recurrent neural network map. 

% Thus, our one can consider this ansatz as a variational Trotterization of evolution $e^{-i\Delta \hat{H}_{\mathrm{eff}}}$ under the Hamiltionian 
% \(H_{\mathrm{eff}} = \sum_q \tfrac{\eta_q}{\lVert\eta_q\rVert_1}\hat{H}_q\)for a time step of size $\Delta =\lVert\eta_q\rVert_1$.

% Though one may construct non-unitary quantum recurrent neural networks, 
% \end{comment}

\subsection{Quantum Graph Convolutional Neural Networks (\textsc{qgcnn})}

Classical Graph Convolutional neural networks rely on a key feature: that of permutation invariance. In other words, the ansatz should be invariant under permutation of the nodes. This is analogous to translational invariance for ordinary convolutional transformations. In our case, permutation invariance manifests itself as a constraint on the Hamiltonian, which now should be devoid of \textit{local} trainable parameters, and should only have global trainable parameters. The $\bm{\theta}$ parameters thus become tied over indices of the graph: $W_{qrjk} \mapsto W_{qr}$ and $B_{qrv}\mapsto B_{qr}$. A broad class of graph convolutional neural networks we will focus on is the set of so-called Quantum Alternating Operator Ansatze~\cite{hadfield2019quantum}, the generalized form of the Quantum Approximate Optimization Algorithm ansatz~\cite{farhi2014quantum}. 

%\textit{Simply tying the parameters across each Hamiltonian such as to obtain permutation invariance/graph independence. We can separate our exponentials that are stricly for message passing (coupling) versus node updating. We consider the \textsc{qaoa} ansatz and most Trotter-based ansatz \cite{} with global variational exponential evolution time parameters to be graph convs. }

%Split into message passing layer, node update, and/or nonlinearity layer. Hamiltonian not with trainable parameters, parameters inherited directly from graph weights in a permutation invariant fashion. This permutation invariance makes it technically a GCN.

\subsection{Quantum Spectral Graph Convolutional Neural Networks (\textsc{qsgcnn})}\label{sec:qsgnn}

We can take inspiration from the continuous-variable quantum approximate optimization ansatz introduced in \cite{verdon2019quantum} to create a variant of the \textsc{qgcnn}: the Quantum Spectral Graph Convolutional Neural Network (\textsc{qsgcnn}). We show here how it recovers the mapping of Laplacian-based graph convolutional networks~\cite{kipf2016semi} in the Heisenberg picture, consisting of alternating layers of message passing, node update, and nonlinearities.

Consider an ansatz of the form from \eqref{eq:QGNN} with four different Hamiltonians ($Q=4$) for a given graph. First, for a weighted graph $\mathcal{G}$ with edge weights $\Lambda_{jk}$, we define the \textit{coupling Hamiltonian} as \[\hat{H}_C  \equiv \tfrac{1}{2}\textstyle\sum_{\{j,k\}\in \mathcal{E}} \Lambda_{jk}(\hat{x}_j - \hat{x}_k)^2.\] The $\Lambda_{jk}$ here are the \textit{weights} of the graph $\mathcal{G}$, and are \textit{not} trainable parameters. The operators denoted here by $\hat{x}_j$ are quantum continuous-variable position operators, which can be implemented via continuous-variable (analog) quantum computers~\cite{weedbrook2012gaussian} or emulated using multiple qubits on digital quantum computers~\cite{somma2015quantum,verdon2018universal}. After evolving by $\hat{H}_C$, which we consider to be the message passing step, one applies an exponential of the \textit{kinetic} Hamiltonian, \( \hat{H}_K \equiv  \tfrac{1}{2}\sum_{j\in \mathcal{V}} \hat{p}_j^2 \). Here $\hat{p}_j$ denotes the continuous-variable momentum (Fourier conjugate) of the position, obeying the canonical commutation relation $[\hat{x}_j,\hat{p}_j]=i\delta_{jk}$. We consider this step as a node update step.  In the Heisenberg picture, the evolution generated by these two steps maps the position operators of each node according to \[e^{-i\gamma \hat{H}_K }e^{-i\alpha \hat{H}_C }:\hat{x}_j\mapsto \hat{x}_j +\gamma\hat{p}_j -\alpha\gamma \textstyle\sum_{k\in \mathcal{V}} L_{jk}\hat{x}_k,\]  where \[L_{jk} =\textstyle \delta_{jk}\left(\sum_{v\in \mathcal{V}} \Lambda_{jv}\right) -\Lambda_{jk} \] is the \textit{Graph Laplacian} matrix for the weighted graph $\mathcal{G}$. We can recognize this step as analogous to classical spectral-based graph convolutions. One difference to note here is that \textit{momentum} is free to accumulate between layers. 

Next, we must add some non-linearity in order to give the ansatz more capacity.\footnote{From a quantum complexity standpoint, adding a nonlinear operation (generated by a potential of degree superior to quadratic) creates states that are \textit{non-Gaussian} and hence are non efficiently simulable on classical computers~\cite{bartlett2002efficient}, in general composing layers of Gaussian and non-Gaussian quantum transformations yields quantum computationally universal ansatz~\cite{lloyd1999quantum}. }
The next evolution is thus generated by an \textit{anharmonic} Hamiltonian \(\hat{H}_A = \sum_{j\in \mathcal{V}} f(\hat{x}_j),\) where $f$ is a nonlinear function of degree greater than 2, e.g., a quartic potential of the form $f(\hat{x}_j) =((\hat{x}_j-\mu)^2 - \omega^2)^2$ for some $\mu, \omega$ hyperparameters. Finally, we apply another evolution according to the kinetic Hamiltonian. These last two steps yield an update \[e^{-i\beta \hat{H}_K }e^{-i\delta \hat{H}_A }:\hat{x}_j\mapsto \hat{x}_j +\beta\hat{p}_j -\delta\beta f'(\hat{x}_j),\] which acts as a nonlinear mapping. By repeating the four evolution steps described above in a sequence of $P$ layers, i.e., 
\[\hat{U}_{\textsc{qsgcnn}}(\bm{\alpha},\bm{\beta},\bm{\gamma}, \bm{\delta}) = \prod_{j=1}^P e^{-i\beta_j \hat{H}_K }e^{-i\delta_j \hat{H}_A }e^{-i\gamma_j \hat{H}_K }e^{-i\alpha_j \hat{H}_C }\]with variational parameters $\bm{\theta} = \{\bm{\alpha},\bm{\beta},\bm{\gamma}, \bm{\delta}\}$,
we then recover a quantum-coherent analogue of the node update prescription of ~\cite{kipf2016semi} in the original graph convolutional networks paper.\footnote{For further physical intuition about the behaviour of this ansatz, note that the sum of the coupling and kinetic Hamiltonians $\hat{H}_K+\hat{H}_C$ is equivalent to the Hamiltonian of a network of quantum harmonic oscillators coupled according to the graph weights and network topology. By adding a quartic $\hat{H}_A$, we are thus emulating parameterized dynamics on a harmonically coupled network of anharmonic oscillators.}

%Can undertand this ansatz as simulating Trotter-like steps of dynamics in a network of harmonic oscillators 

\section{Applications \& Experiments}

\subsection{Learning Quantum Hamiltonian Dynamics with Quantum Graph Recurrent Neural Networks}\label{sec:dynamics}

\begin{figure}[t!]
    \centering
    \includegraphics[width=0.55\textwidth]{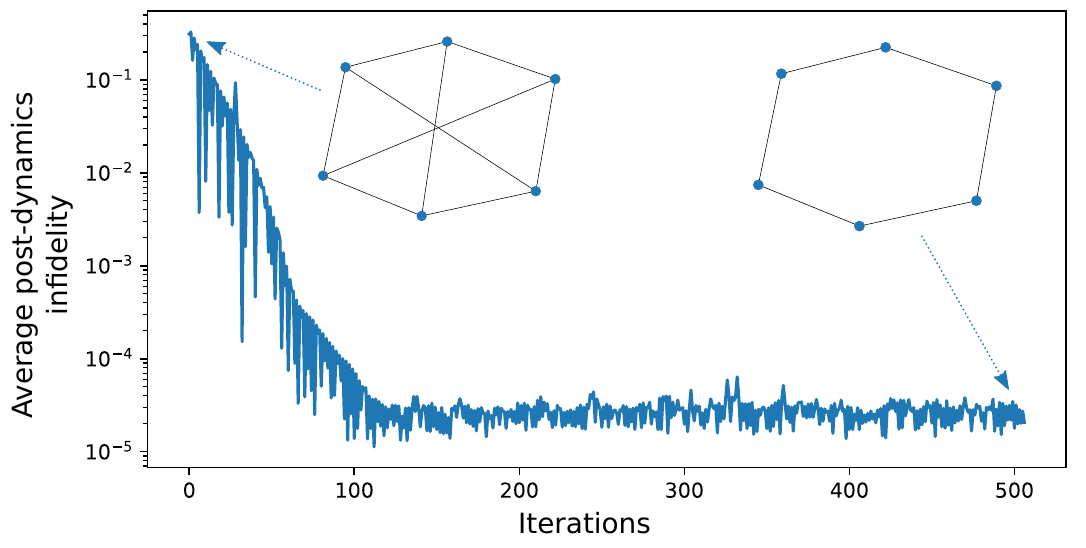}\includegraphics[width=0.25\textwidth]{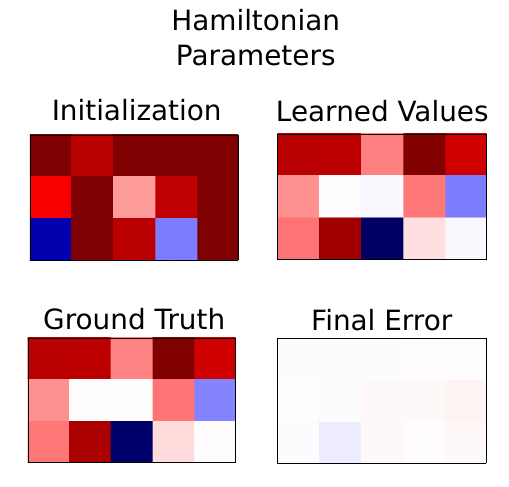}
  \caption{Left: Batch average infidelity with respect to ground truth state sampled at 15 randomly chosen times of quantum Hamiltonian evolution. We see the initial guess has a densely connected topology and the \textsc{qgrnn} learns the ring structure of the true Hamiltonian. Right: Ising Hamiltonian parameters (weights \& biases) on a color scale.}
  \label{fig:qgrnn}
\end{figure}
Learning the dynamics of a closed quantum system is a task of interest for many applications \cite{wiebe2014hamiltonian}, including device characterization and validation. In this example, we demonstrate that a Quantum Graph Recurrent Neural Network can learn effective dynamics of an Ising spin system when given access to the output of quantum dynamics at various times.

Our target is an Ising Hamiltonian with transverse field on a particular graph, \[\textstyle\hat{H}_{\text{target}} = \sum_{\{j,k\}\in \mathcal{E}} J_{jk}\hat{Z}_j\hat{Z}_k + \sum_{v\in\mathcal{V}}Q_v\hat{Z}_v + \sum_{v\in\mathcal{V}} \hat{X}_j.\] We are given copies of a fixed low-energy state  $\ket{\psi_0}$ as well as copies of the state $\ket{\psi_T} \equiv \hat{U}(T)\ket{\psi_0} = e^{-i T \hat{H}_{\text{target}}}$ for some known but randomly chosen times $T\in[0,T_{\text{max}}]$. Our goal is to learn the target Hamiltonian parameters $\{J_{jk}, Q_v\}_{j,k,v\in\mathcal{V}}$ by comparing the state $\ket{\psi_T}$ with the state obtained by evolving $\ket{\psi_0}$ according to the \textsc{qgrnn} ansatz for a number of iterations $P \approx T/\Delta$ (where $\Delta$ is a hyperparameter determining the Trotter step size). We achieve this by training the parameters via Adam~\cite{kingma2014adam} gradient descent on the average infidelity $\mathcal{L}(\bm{\theta}) =  1-\tfrac{1}{B}\sum_{j=1}^B  |\braket{\psi_{T_j}|U^j_{\textsc{qgrnn}}( \Delta,\bm{\theta})\ket{\psi_0}}|^2$ averaged over batch sizes of 15 different times $T$. Gradients were estimated via finite difference differentiation with step size $\epsilon =10^{-4}$. The fidelities (quantum state overlap) between the output of our ansatz and the time-evolved data state were estimated via the quantum swap test \cite{cincio2018learning}. The ansatz uses a Trotterization of a random densely-connected Ising Hamiltonian with transverse field as its initial guess, and successfully learns the Hamiltonian parameters within a high degree of accuracy as shown in Figure~1a.

% \begin{wrapfigure}{l}{0.5\textwidth}
%   \begin{center}
%     \includegraphics[width=0.25\textwidth]{GHZ_loss.pdf}\includegraphics[width=0.25\textwidth]{GHZ_phase.pdf}
%   \caption{Left: Stabilizer Hamiltonian expectation and fidelity over training iterations. A picture of the quantum network topology is inset. Right: Quantum phase kickback test on the learned GHZ state. We observe a $7$x boost in Rabi oscillation frequency for a 7-node network, thus demonstrating we have reached the Heisenberg limit of sensitivity for the quantum sensor network.}
%   \label{fig:ghz}
%     \end{center}
% \end{wrapfigure}

\subsection{Quantum Graph Convolutional Neural Networks for Quantum Sensor Networks}

\begin{figure}[t!]
  \centering
    \includegraphics[width=0.4\textwidth]{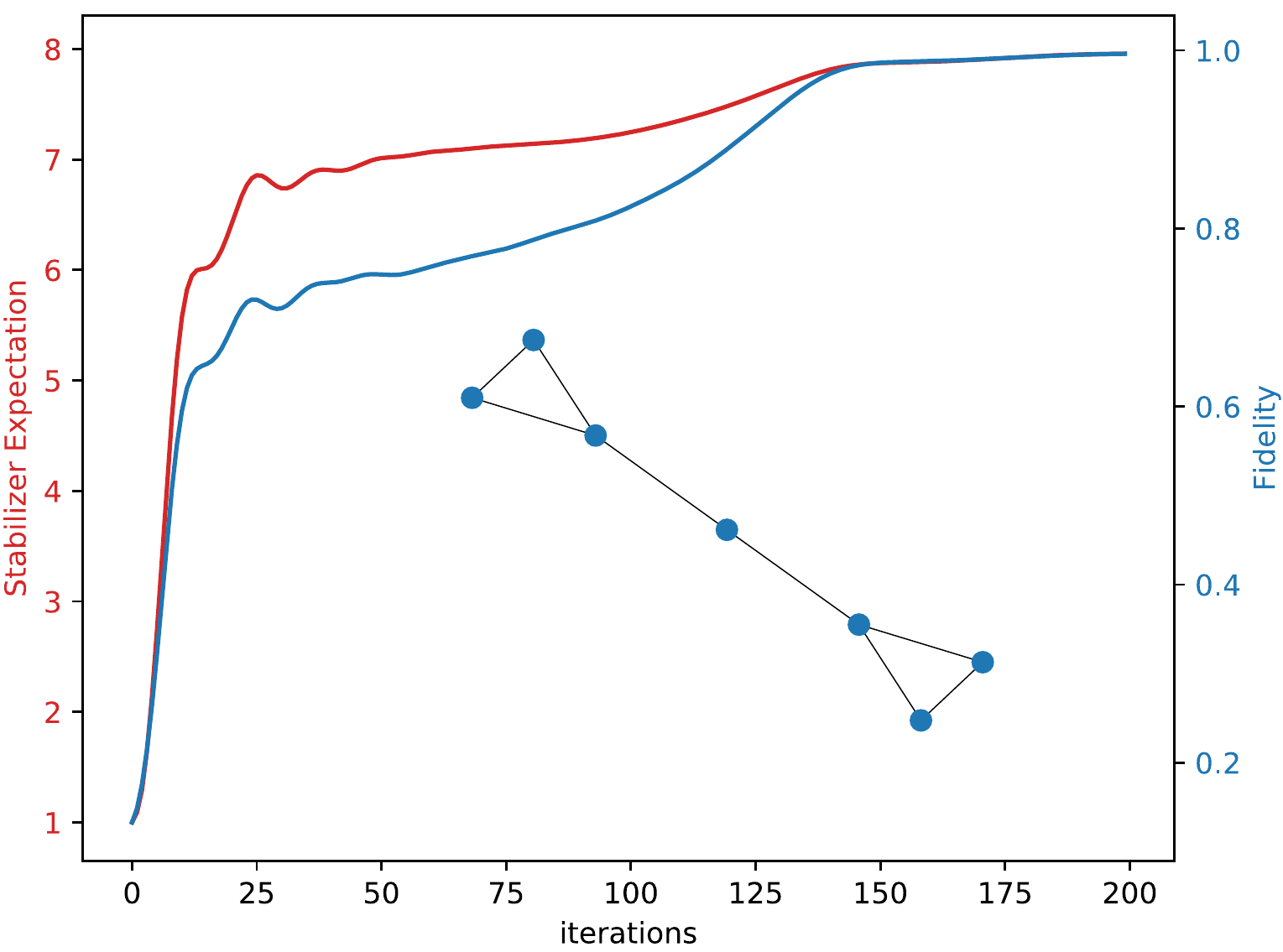}\hspace{2em}\includegraphics[width=0.4\textwidth]{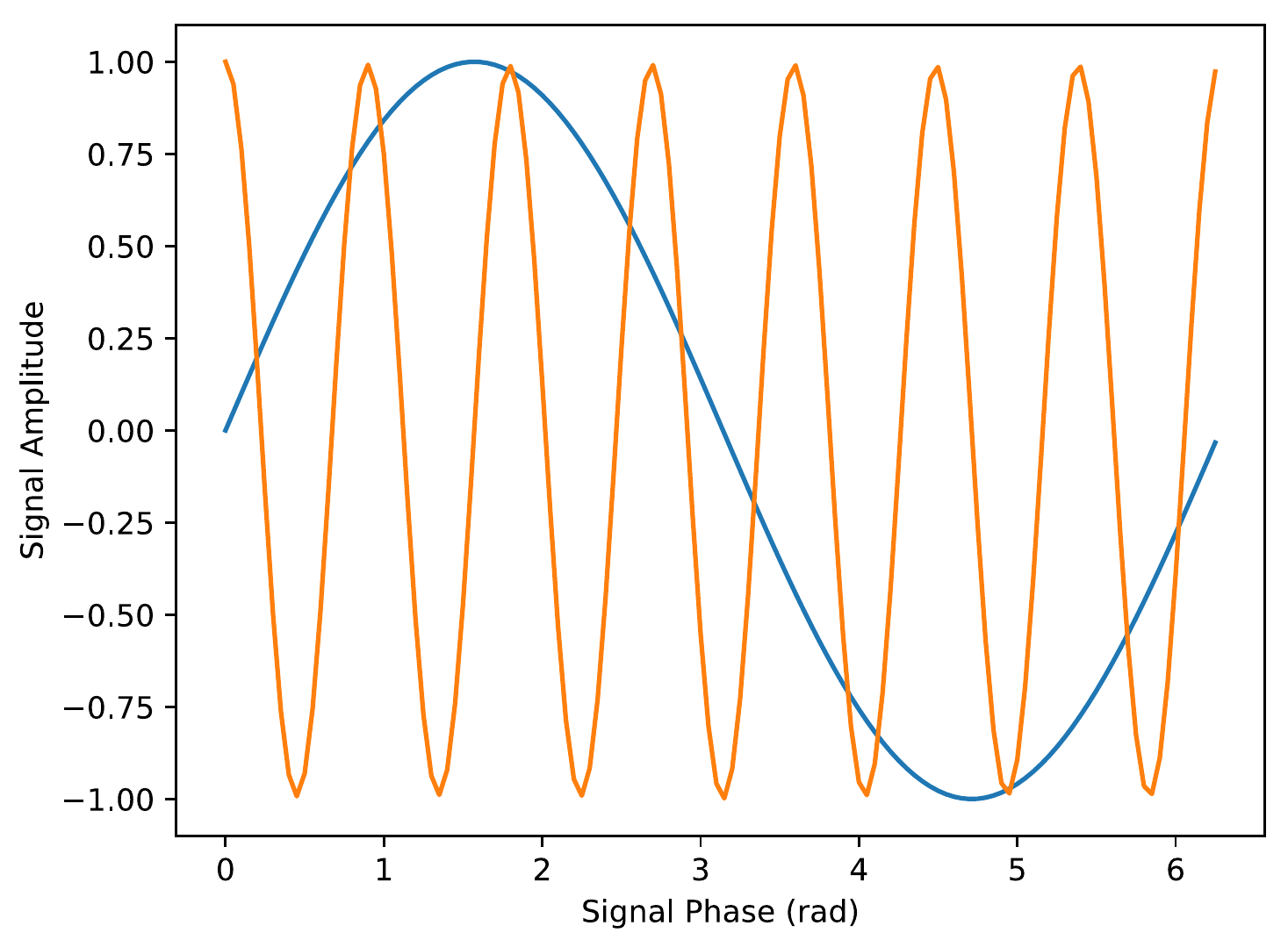}
  \caption{Left: Stabilizer Hamiltonian expectation and fidelity over training iterations. A picture of the quantum network topology is inset. Right: Quantum phase kickback test on the learned GHZ state. We observe a $7$x boost in Rabi oscillation frequency for a 7-node network, thus demonstrating we have reached the Heisenberg limit of sensitivity for the quantum sensor network.}
  \label{fig:ghz}
\end{figure}
Quantum Sensor Networks are a promising area of application for the technologies of Quantum Sensing and Quantum Networking/Communication \cite{kimble2008quantum,qian2019heisenberg}. A common task considered where a quantum advantage can be demonstrated is the estimation of a parameter hidden in weak qubit phase rotation signals, such as those encountered when artificial atoms interact with a constant electric field of small amplitude \cite{qian2019heisenberg}. A well-known method to achieve this advantange is via the use of a quantum state exhibiting multipartite entanglement of the Greenberger–Horne–Zeilinger kind, also known as a GHZ state~\cite{greenberger1989going}. Here we demonstrate that, without global knowledge of the quantum network structure, a \textsc{qgcnn} ansatz can learn to prepare a GHZ state. We use a \textsc{qgcnn} ansatz with $\hat{H}_1 = \sum_{\{j,k\}\in\mathcal{E}} \hat{Z}_j\hat{Z}_k$ and $\hat{H}_2= \sum_{j\in\mathcal{V}} \hat{X}_j$. The loss function is the negative expectation of the sum of stabilizer group generators which stabilize the GHZ state \cite{toth2005entanglement}, i.e., \[\mathcal{L}(\bm {\eta}) = -\braket{\textstyle \bigotimes_{j=0}^n\hat{X} +\sum_{j=1}^{n-1}\hat{Z}_j\hat{Z}_{j
+1}}_{\bm{\eta}}\] for a network of $n$ qubits. Results are presented in Figure~1b. Note that the advantage of using a \textsc{qgnn} ansatz on the network is that the number of quantum communication rounds is simply proportional to $P$, and that the local dynamics of each node are independent of the global network structure. 

In order to further validate that we have obtained an accurate GHZ state on the network after training, we perform the quantum phase kickback test on the network's prepared approximate GHZ state \cite{wei2019verifying}.\footnote{For this test, one applies a phase rotation $\bigotimes_{j\in\mathcal{V}} e^{-i\varphi \hat{Z}_j}$ on all the qubits in paralel, then one applies a sequence of CNOTs (quantum adder gates) to concentrate the phase shifts onto a single collector node, $m\in\mathcal{V}$. Given that one had a GHZ state initially, one should then observe a phase shift $e^{-in\varphi \hat{Z}_m}$ where $n = |\mathcal{V}|$. This boost in frequency of oscillation of the signal is what gives quantum multipartite entanglement its power to increase sensitivity to signals to super-classical levels \cite{degen2017quantum}.} We observe the desired frequency boost effect for our trained network preparing an approximate GHZ state at test time, as displayed in Figure~\ref{fig:ghz}.

\subsection{Unsupervised Graph Clustering with Quantum Graph Convolutional Networks}
\begin{figure}%{r}{0.5\textwidth}
  \begin{center}
    \includegraphics[width=0.63\linewidth]{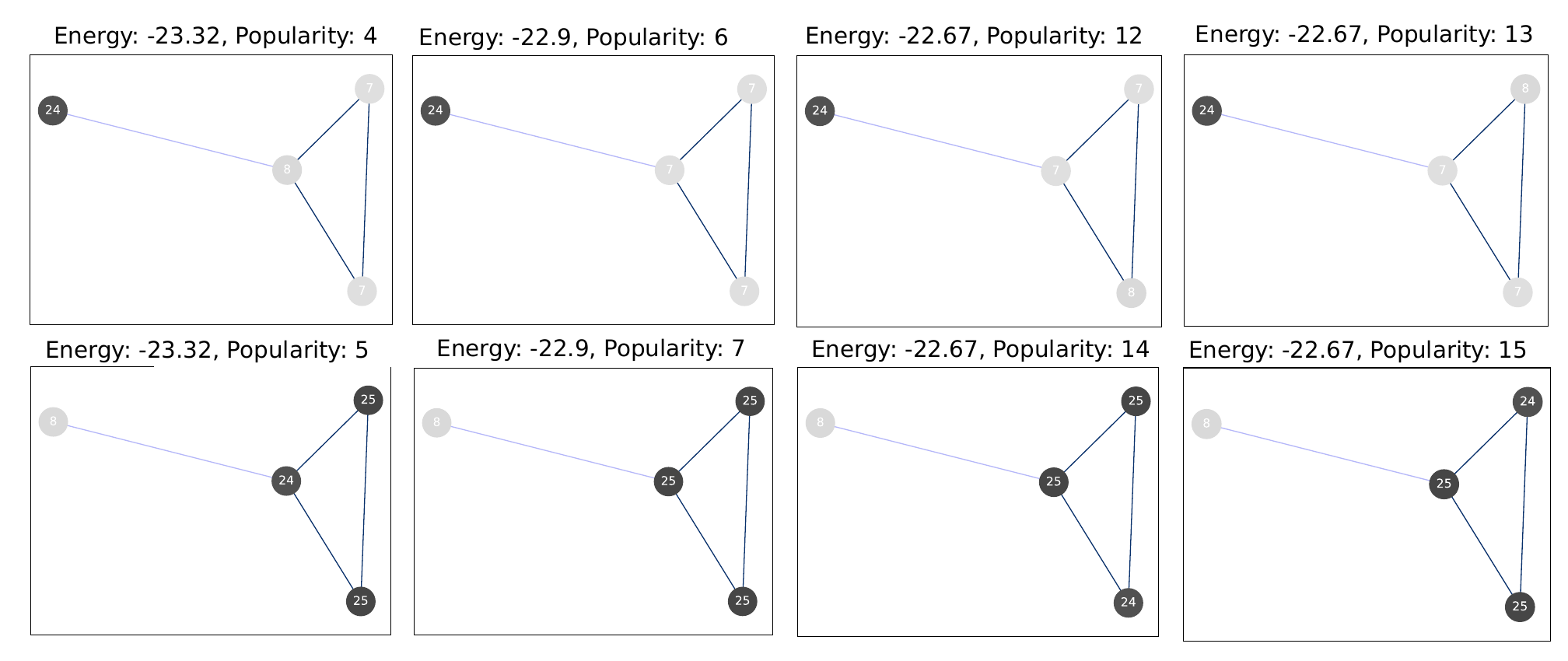}\includegraphics[width=0.35\linewidth]{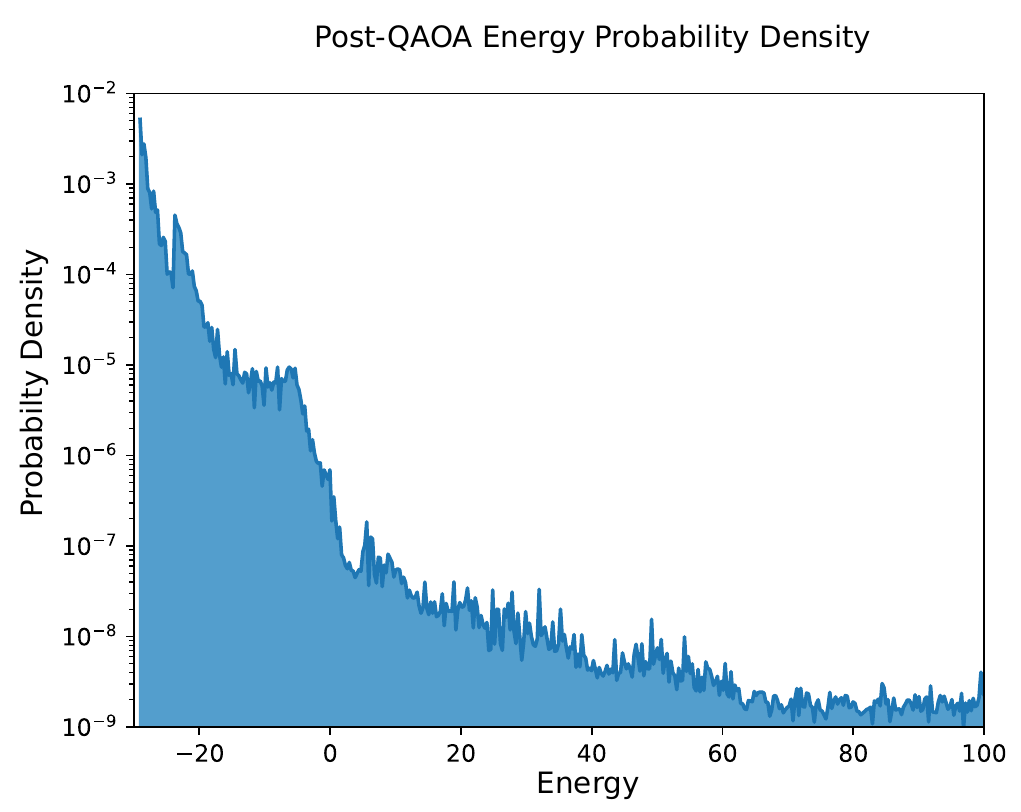}
    
    \includegraphics[width=0.62\linewidth]{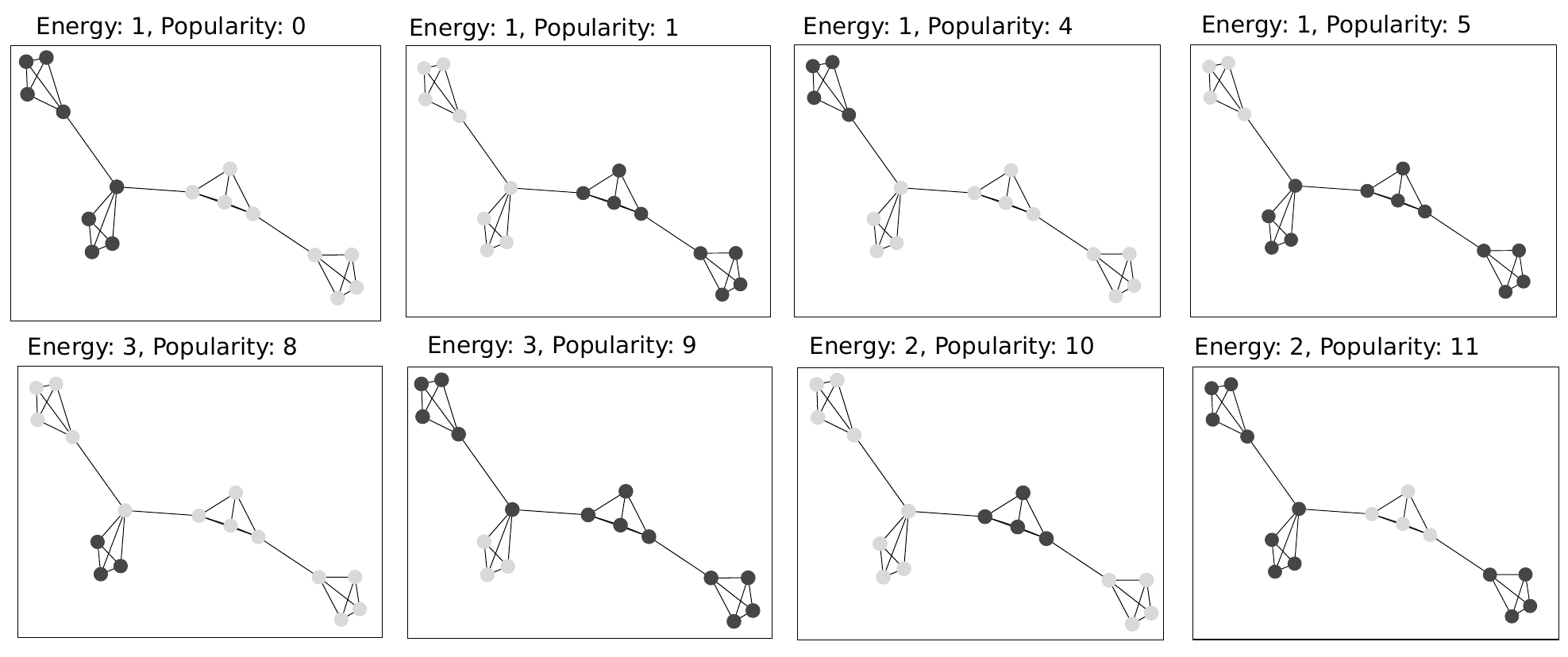}\includegraphics[width=0.35\linewidth]{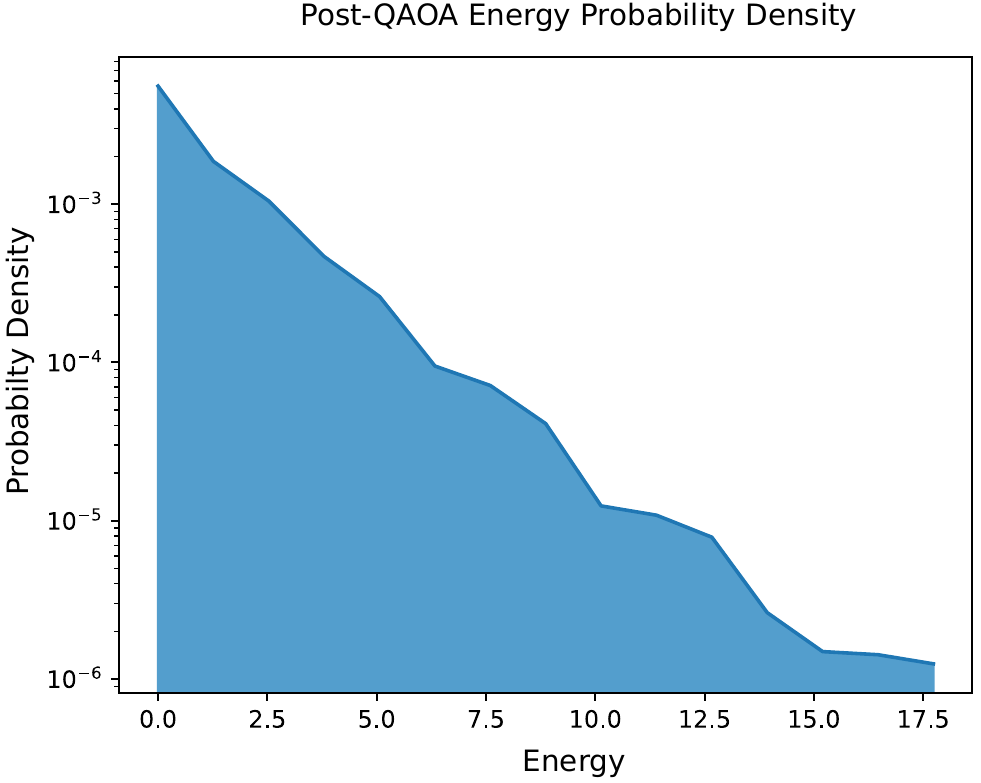}

  \caption{\textsc{qsgcnn} spectral clustering results for  5-qubit precision (top) with quartic double-well potential and 1-qubit precision (bottom) for different graphs. Weight values are represented as opacity of edges, output sampled node values as grayscale.  Lower precision allows for more nodes in the simulation of the quantum neural network. The graphs displayed are the most probable (populated) configurations, and to their right is the output probability distribution over potential energies. We see lower energies are most probable and that these configurations have node values clustered.}
  \label{fig:spectral}
    \end{center}
\end{figure}

As a third set of applications, we consider applying the \textsc{qsgcnn} from Section~\ref{sec:qsgnn} to the task of spectral clustering \cite{ng2002spectral}. Spectral clustering involves finding low-frequency eigenvalues of the graph Laplacian and clustering the node values in order to identify graph clusters. In Figure~\ref{fig:spectral} we present the results for a \textsc{qsgcnn} for varying multi-qubit precision for the representation of the continuous values, where the loss function that was minimized was the expected value of the anharmonic potential $\mathcal{L}(\bm{\eta}) = \braket{\hat{H}_C + \hat{H}_A}_{\bm{\eta}}$. Of particular interest to near-term quantum computing with low numbers if qubits is the single-qubit precision case, where we modify the \textsc{qsgcnn} construction as $\hat{p}^2_j\mapsto \hat{X}_j$,  $\hat{H}_A\mapsto I $ and $\hat{x}_j\mapsto \ket{1}\!\bra{1}_j$ which transforms the coupling Hamiltonian as 
\begin{equation}\label{eq:1qb_cost} \hat{H}_C \mapsto  \tfrac{1}{2}\textstyle\sum_{\{j,k\}\in \mathcal{E}} \Lambda_{jk}(\ket{1}\!\bra{1}_j - \ket{1}\!\bra{1}_k)^2 = \sum_{jk} L_{jk} \ket{1}\!\bra{1}_j\otimes \ket{1}\!\bra{1}_k,\end{equation}
where $\ket{1}\!\bra{1}_k = (\hat{I}-\hat{Z}_k)/2$. We see that using a low-qubit precision yields sensible results, thus implying that spectral clustering could be a promising new application for near-term quantum devices.

\vspace{1em}

\subsection{Graph Isomorphism Classification via Quantum Graph Convolutional Networks}

Recently, a benchmark of the representation power of classical graph neural networks has been proposed \cite{xu2018powerful} where one uses classical \textsc{gcn}s to identify whether two graphs are isomorphic. In this spirit, using the \textsc{qsgcnn} ansatz from the previous subsection, we benchmarked the performance of this Quantum Graph Convolutional Network for identifying isomorphic graphs. We used the single-qubit precision encoding in order to order to simulate the execution of the quantum algorithms on larger graphs. 

Our approach was the following, given two graphs $\mathcal{G}_1$ and $\mathcal{G}_2$, one applies the single-qubit precision \textsc{qsgcnn} ansatz \(\prod_{j=1}^P e^{i\eta_j \hat{H}_K}e^{i\gamma_j \hat{H}_C}\) with $\hat{H}_K= \sum_{j\in\mathcal{V}} \hat{X}_j$ and $\hat{H}_C$ from \eqref{eq:1qb_cost} in parallel according to each graph's structure. One then samples eigenvalues of the coupling Hamiltonian $\hat{H}_C$ on both graphs via standard basis measurement of the qubits and computation of the eigenvalue at each sample of the wavefunction. One then obtains a set of samples of ``energies'' of this Hamiltonian. By comparing the energetic measurement statistics output by the  \textsc{qsgcnn} ansatz applied with identical parameters $\bm{\theta} = \{\bm{\eta},\bm{\gamma}\}$ for two different graphs, one can then infer whether the graphs are isomorphic.

We used the Kolmogorov-Smirnoff test \cite{lilliefors1967kolmogorov} on the distribution of energies sampled at the output of the \textsc{qsgcnn} to determine whether two given graphs were isomorphic. In order to determine the binary classification label deterministically, we considered all KS statistic values above $0.4$ to indicate that the graphs were non-isomorphic. For training and testing purposes, we set the loss function to be $\mathcal{L}(y, \ks) = (1-y)(1-\ks) + y \ks$, where $y=1$ if graphs are isomorphic, and $y=0$ otherwise.

For the dataset, graphs were sampled uniformly at random from the Erdos-Renyi distribution $G(n,p)$ with $p=0.5$ at fixed $n$. In all of our experiments, we had 100 pairs of graphs for training, 50 for validation, and 50 for testing, always balanced between isomorphic and non-isomorphic pairs. Moreover, we only considered graphs that were connected. The networks were trained via a Nelder-Mead optimization algorithm.

\begin{figure}[t!]
  \centering
    \includegraphics[width=0.5\textwidth]{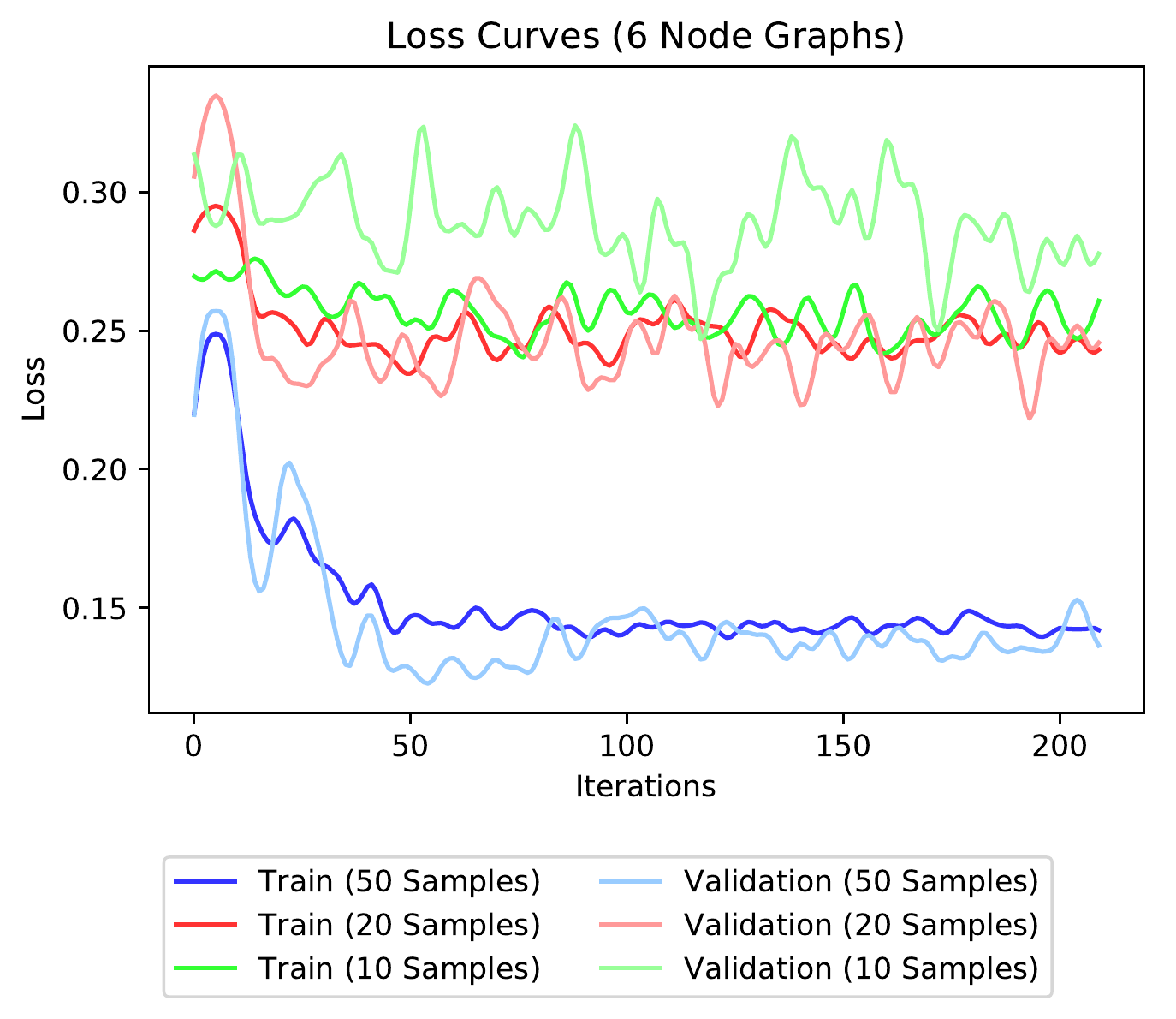}\includegraphics[width=0.5\textwidth]{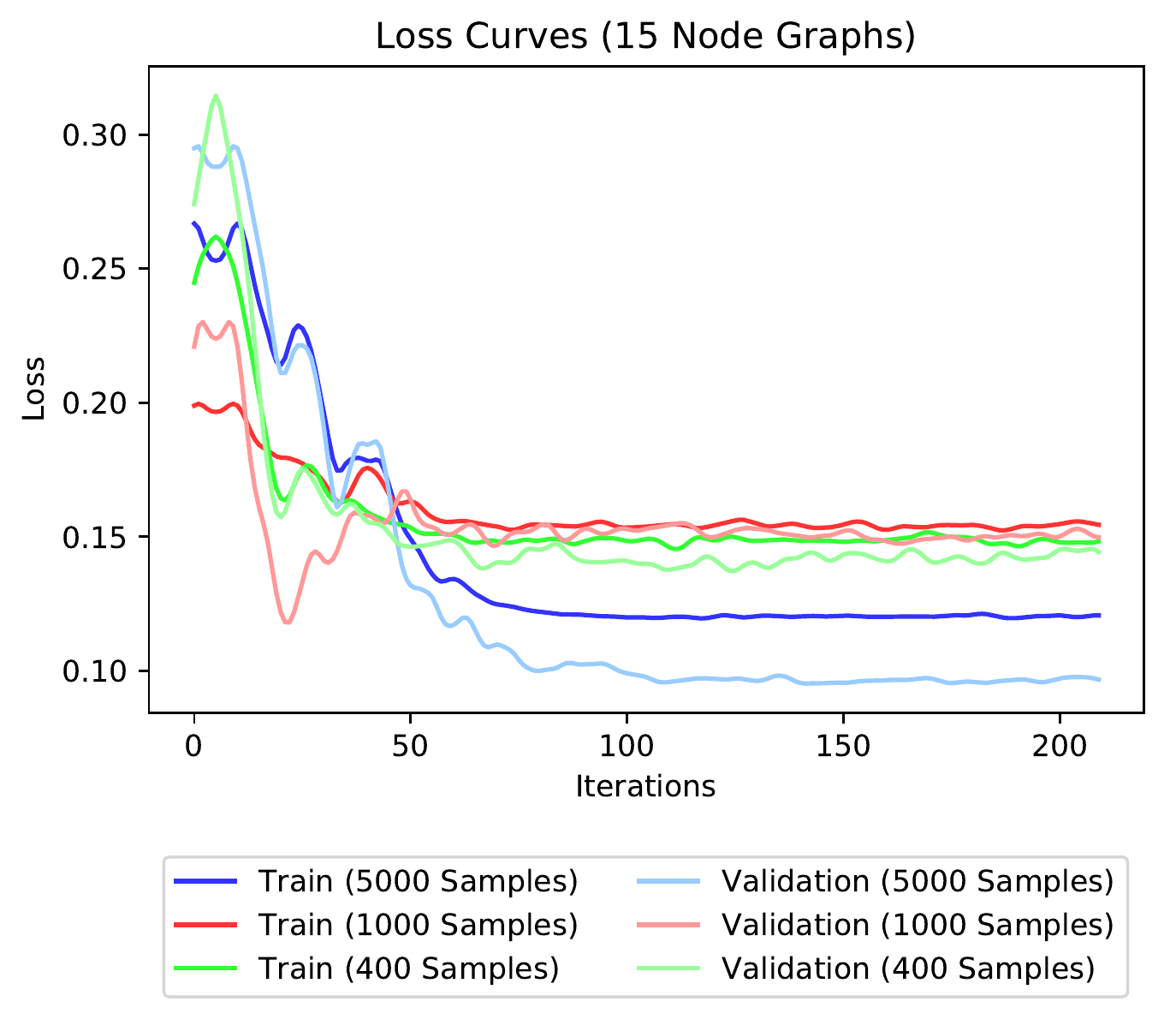}
  \caption{Graph isomorphism loss curves for training and validation for various numbers of samples. Left is for 6 node graphs and right is for 15 node graphs. The loss is based on the Kolmogorov-Smirnov statistic comparing the sampled distribution of energies of the \textsc{qgcnn} output on two graphs. }
  \label{fig:iso_loss}
\end{figure}

Presented in Figure~\ref{fig:iso_loss} is the training and testing losses for various graph sizes and numbers of energetic samples. In Tables 1 and 2, we present the graph isomorphism classification accuracy for the training and testing sets using the trained \textsc{qgcnn} with the previously described thresholded KS statistic as the label. We see we get highly accurate performance even at low sample sizes. This seems to imply that the  \textsc{qgcnn} is fully capable of identifying graph isomorphism, as desired for graph convolutional network benchmarks.

 %From what distribution did you sample the graphs 

\begin{table}[]
\label{sample-table}
\begin{center}
\vspace{2.5em}
\parbox{.45\linewidth}{
\caption{Classification Accuracy for 15 Node Graphs}
\begin{tabular}{rrr}
\multicolumn{1}{c}{\bf  Samples}  &\multicolumn{1}{c}{\bf Test}  &\multicolumn{1}{c}{\bf Validation}
\\ \hline
5000         &100.0 &100.0\\
1000             &100.0 &100.0 \\
400             &100.0 &100.0 \\
\end{tabular}
}
\quad
\parbox{.45\linewidth}{
\caption{Classification Accuracy for 6 Node Graphs}
\label{sample-table}
\begin{tabular}{rrr}
\multicolumn{1}{c}{\bf Samples}  &\multicolumn{1}{c}{\bf Test }  &\multicolumn{1}{c}{\bf Validation }
\\ \hline 
50         &100.0  &100.0\\
20             &90.0 &100.0\\
10             &100.0  &80.0\\
\end{tabular}
}
\end{center}
\end{table}

\section{Conclusion \& Outlook}

Results featured in this paper should be viewed as a promising set of first explorations of the potential applications of \textsc{qgnn}s. Through our numerical experiments, we have shown the use of these \textsc{qgnn} ansatze in the context of quantum dynamics learning, quantum sensor network optimization, unsupervised graph clustering, and supervised graph isomorphism classification. Given that there is a vast set of literature on the use of Graph Neural Networks and their variants to quantum chemistry, future works should explore hybrid methods where one can learn a graph-based hidden quantum representation (via a \textsc{qgnn}) of a quantum chemical process. As the true underlying process is quantum in nature and has a natural molecular graph geometry, the \textsc{qgnn} could serve as a more accurate model for the hidden processes which lead to perceived emergent chemical properties. We seek to explore this in future work. Other future work could include generalizing the \textsc{qgnn} to include quantum degrees of freedom on the edges, include quantum-optimization-based training of the graph parameters via quantum phase backpropagation \cite{verdon2018universal}, and extending the \textsc{qsgcnn} to multiple features per node.

\subsubsection*{Acknowledgments}
Numerics in this paper were executed using a custom interface between Google's Cirq~\cite{cirq} and TensorFlow~\cite{abadi2016tensorflow}. The authors would like to thank Edward Farhi, Jae Yoo, and Li Li for useful discussions. GV, EL, and VS would like to thank the team at X for the hospitality and support during their respective Quantum@X and AI@X residencies where this work was completed. X, formerly known as Google[x], is part of the Alphabet family of companies, which includes Google, Verily, Waymo, and others (www.x.company). GV acknowledges funding from NSERC. 

\bibliography{references}
\bibliographystyle{unsrt}

\end{document}